\documentclass[aps,pre,showpacs]{revtex4-1}
\usepackage{amsmath}
\usepackage{amssymb}
\usepackage{amsfonts}
\usepackage{makeidx}
\usepackage{graphicx}
\include{epsf}
\usepackage{bm}
\usepackage{clrscode}
\usepackage{color}
\usepackage{subcaption}

\newcommand{\brc}[1]{\left\langle#1\right\rangle}

\newcommand{\bea}{\begin{eqnarray}}
\newcommand{\eea}{\end{eqnarray}}

\def\be{\begin{equation}}
\def\ee{\end{equation}}

\begin{document} 

\title{Inference of the sparse kinetic Ising model using the decimation method}

\author{Aur\'elien Decelle$^{1,2}$ and Pan Zhang$^{3,4}$}

\affiliation{
$^1$Dipartimento di Fisica, Universit\`a La Sapienza, Piazzale Aldo Moro 5, I-00185 Roma, Italy. \\
$^2$Laboratoire de Recherche en Informatique, TAO - INRIA, CNRS et Universit\'e Paris-Sud, B\^at. 660, 91190 Gif-sur-Yvette, France\\
$^3$Santa Fe Institute, 1399 Hyde Park Road, Santa Fe, New Mexico 87501, USA  \\
$^4$State Key Laboratory of Theoretical Physics, Institute of Theoretical Physics, Chinese Academy of Sciences, Beijing 100190, China}


\begin{abstract}
	In this paper we study the inference of the kinetic Ising model on sparse graphs by the decimation method.
	The decimation method, which was first proposed in 
	[Phys. Rev. Lett. 112, 070603] for the static inverse Ising problem, tries to recover the topology of the inferred system by setting the weakest couplings to zero iteratively. During the decimation process the likelihood function is maximized over the remaining couplings.
	Unlike the $\ell_1$-optimization based methods, the decimation method 
	does not use
	the Laplace distribution as a heuristic choice of prior to select a sparse solution.
	In our case, the whole process can be done automatically without fixing any parameters by hand.
	We show that in the dynamical inference problem, where the task is to reconstruct
	the couplings of an Ising model given the data, the 
	decimation process can be applied naturally into a maximum-likelihood optimization algorithm, 
	as opposed to the static case where pseudo-likelihood method needs to be adopted. 
We also use extensive numerical studies to validate the accuracy of our methods in 
dynamical inference problems. 
Our results illustrate that on various topologies and with different 
distribution of couplings, the decimation method outperforms 
the widely-used $\ell _1$-optimization based methods.
\end{abstract}
\maketitle

\section{Introduction}\label{sec:intro}
In many fields of science, a large amount of effort has been devoted to 
theories and methods to do inference from observed data. Recently, 
there has been a considerable attention drawn to the inverse Ising problem in performing
such task. This problem, also known as ``Boltzmann 
machine learning'' \cite{ackley1985learning}, focused on finding the parameters of an Ising
model according to data that are sampled from the Boltzmann distribution of the
original system. The interest in this particular model
is linked to the maximum entropy principle applied to pairwise interacting variables 
when the first two moments are measurable \cite{Schneidman_etal_2006_nature}. It is 
also applied to a large number of relevant datasets coming from many different fields.
It has been used not only in physics \cite{aurell2012inverse,inf_nguyen-11,inf_nguyen-12,ricci2011mean,decelle13} and
computer science \cite{wainwright2010AnnalHigh}, but also in biology
(gene networks \cite{ruz2010learning} and protein folding \cite{weigtPNAS,ekeberg2013improved}), neuroscience
 \cite{Schneidman_etal_2006_nature, tkacik2006ising}, social network \cite{fortunato2010community} and statistics of birds flock \cite{bialek2012statistical}. 
The fundamental approach to inference problems in general is based on
maximizing the likelihood function of the model. This function
represents the probability of generating the data given the parameters
of the underlying model.

\begin{figure}[!h]
\centering
   \begin{subfigure}[b]{0.325\textwidth}
	      \includegraphics[width=1\columnwidth]{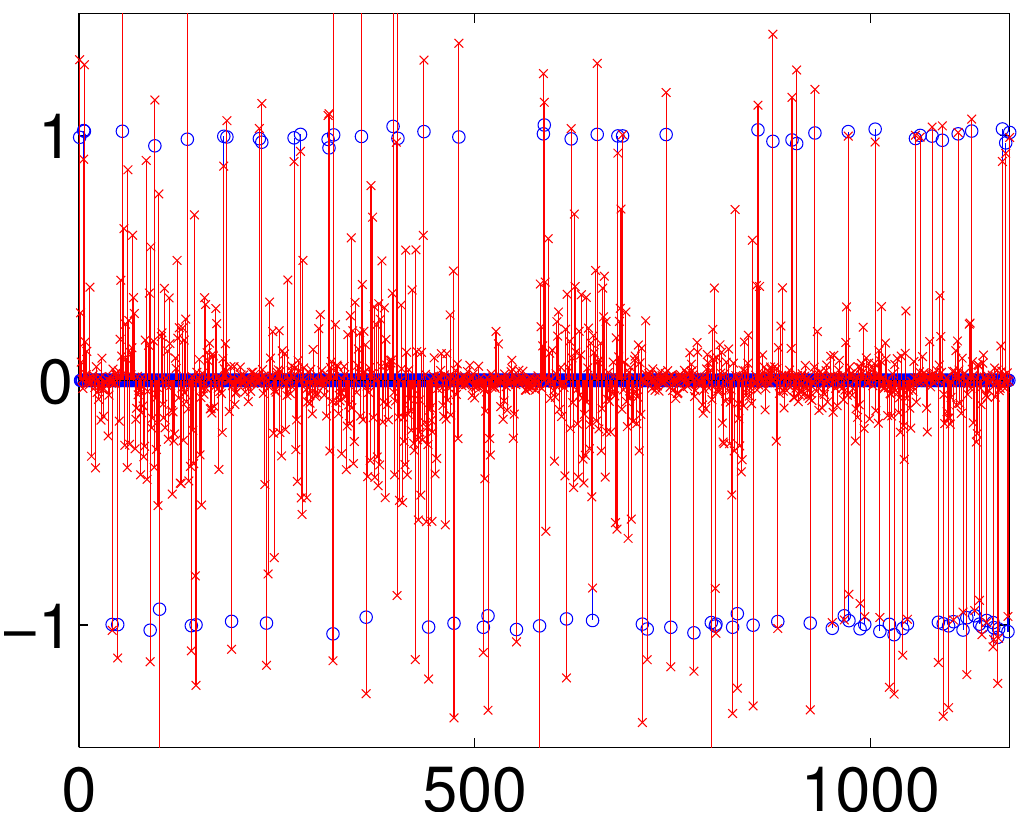}
		  \caption{}
	  \end{subfigure}
   \begin{subfigure}[b]{0.325\textwidth}
	      \includegraphics[width=1\columnwidth]{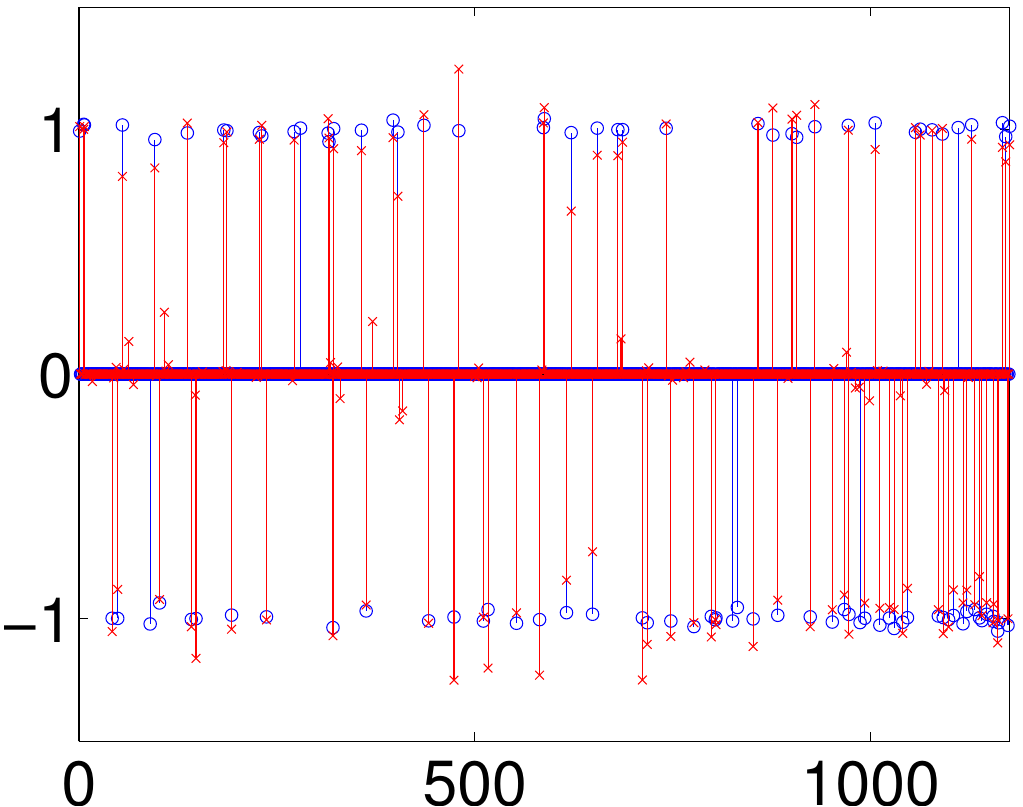}
		  \caption{}
	  \end{subfigure}
   \begin{subfigure}[b]{0.325\textwidth}
	      \includegraphics[width=1\columnwidth]{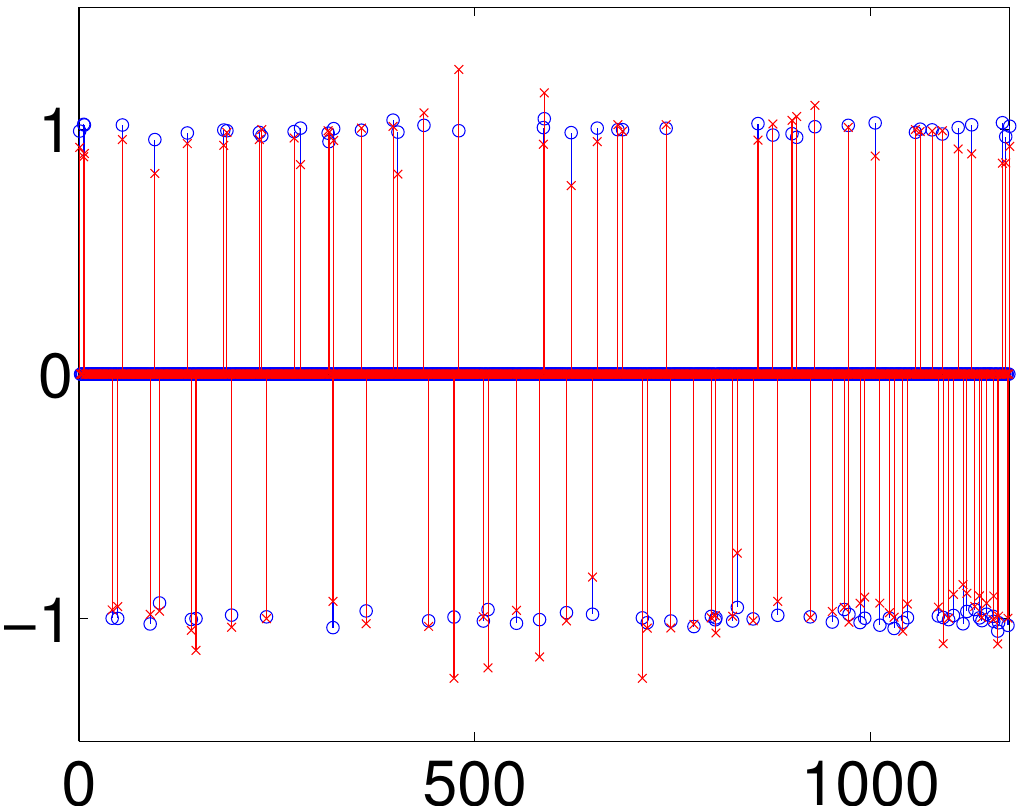}
		  \caption{}
	  \end{subfigure}
  \caption{(Color online) True couplings (blue circles) compared with couplings inferred (red crosses) 
  by maximizing likelihood method (a), $\ell_1$ minimization + refinement (b) and 
  Decimation (c). The relative error (Eq.~\eqref{eq:diff}) is $0.209$, $0.079$ and $0.026$ from left to right respectively.
   The network is a $2$D lattice with size $N=49$, $\beta=2.0$, $R=10$ and $T=500$.}
  \label{fig:maxlike}
\end{figure}

The inferred parameters are therefore the ones who maximize the
likelihood. However, approaches based on the maximization of the
likelihood function are usually prone to overfitting, tending to fit
not only the experimental data but also the noise.  Over-fitting
is a common problem in Bayesian inference and it is crucial to find
methods that minimize this effect in order to obtain a model that can
describe correctly real experiments. Moreover, a lot of real systems
in which we are interested, e.g. biological systems and social
systems, are sparse: the topology (or network) of the system has many
empty interactions (edges).  In those sparse systems, the
over-fitting problem turns out to prevent the inference process to
reconstruct properly the topology of the system.  Therefore, making
progress on the inference process, both on the complexity of the
algorithm and on the accuracy of the inference, could not only allow
us to deal with larger system sizes, but also to retrieve essential
information of the considered system; for instance which edges are
absent. Thus, in the context of the inverse Ising problem, the over-fitting
induced by maximizing the likelihood does not only increase the
error made on the inferred couplings, but it also prevents from recovering the
proper topology of the system.  As an example, we illustrate in the left
panel of Fig.~\ref{fig:maxlike} the true couplings of a sparse Ising
model with the couplings inferred by maximizing the likelihood.  From the
figure we can see that, although the original network has only few
large couplings (non-existing couplings being zero), maximizing the
likelihood gives barely zero-valued edges, thus ending with a
completely wrong topology of the network.

In this article, we focus on the important problem of finding the
topology of the underlying problem from the observed data or, in
another words, we want to separate the most important interactions
with respect to the null ones. We believe that the progress made
toward this direction in the context of the inverse Ising problem will
be fruitful when applied to the inference of real systems.

Various approaches \cite{wainwright2010AnnalHigh, cocco2011adaptive,
  aurell2012inverse, CoccoPNAS, decelle13} have been proposed for the
sparse inverse Ising problem, with different ways to overcome the
over-fitting problem. Among various approaches, the probably most
famous and effective one is based on the $\ell_1$-regularization
\cite{wainwright2010AnnalHigh} which consists in adding a prior on
the couplings. This prior takes the form of a Laplace distribution on
the parameters and usually leads to a sparse solution.  As an example,
in the middle panel of Fig.\ref{fig:maxlike} we compared the true
couplings with the couplings inferred by the $\ell_1$ minimization
approach on a sparse system.  From the figure, we can
see that the $\ell_1$ approach indeed recover much better the null couplings and
give a smaller error than maximizing the likelihood alone.  However, in our
experiments, the $\ell_1$ approach still does not work perfectly, and
fails to recover correctly the topology in many cases (even some easy
ones).  Moreover, its performance depends on an external parameter
$\lambda$ that has to be chosen heuristically.

Recently, a decimation process was proposed in \cite{decelle13} for
inferring the topology in the static inverse Ising problem. It was
shown to give a large improvement over the reconstructed topology of
the inference process over $\ell_1$-based methods.  In the static
problem, empirical data are equilibrium configurations that are sampled
from the Boltzmann distribution $\mathcal
P(\underline{\sigma})=\frac{1}{Z}e^{-\beta E(\underline{\sigma})}$,
where $Z$ is the partition function and
$E(\underline{\sigma})=-\sum_{\brc{i,j}}J_{ij}\sigma_i\sigma_j$
denotes the energy of a configuration $\underline{\sigma}$. When the
number of samples is large, the couplings can be determined by
maximizing the likelihood $\mathcal
L(\underline{J})=-\brc{E(\underline{J})}_D-\log Z(\underline{J})$,
where $\brc{E}_D$ denotes the averaged value of the energy over the
data. In this setting however, the partition function $Z$ is hard to
compute exactly to compute exactly and therefore the likelihood
function is hard to compute as well (or to maximize).  In
\cite{decelle13,yamanaka2015detection}, in order to use the decimation method, the authors
needed to adopt the ``pseudo-likelihood'' which is essentially an
approximation of the true likelihood function
\cite{besag1975statistical,ravikumar2010high}.

In this work, we deal with the dynamical inverse Ising problem. This
type of inference problem introduces a new sets of behavior that are
absent of the static counter-part. Indeed, the configurations are now
drawn from a stochastic process and are therefore correlated in
time. From this stochastic process we expect the presence of relaxation
toward an equilibrium distribution (somehow similar to the static
inverse Ising case) but such systems also exhibit ergodicity breaking
or limited cycles. In addition, it is more natural to describe many real
systems by a dynamical process rather than by using an equilibrium
distribution. For instance some biological systems depends on a
time-dependent external stimuli. Moreover and as opposed to the static case, 
the couplings between the variables need not to be symmetric. Again, we show on
the Fig. \ref{fig:maxlike} right panel an illustration of this algorithm on the same system
as before (see the caption) and we can see it manages to better recover
the topology of the network than maximizing the likelihood alone.


The rest of the paper is organized as follows. Section \ref{sec:model} includes
definitions and the description of the dynamical inverse Ising problem.
In Sec. \ref{sec:sparse} we consider the dynamical inverse Ising
problem on sparse graphs and review the popular $\ell_1$ based
approach. In Sec. \ref{sec:dec} we study the inference of the
couplings using the decimation process.  In Sec. \ref{sec:result} we
compare our method with the $\ell_1$ approach on a larger set of examples.  We conclude this work
in Sec. \ref{sec:con}.

\section{The dynamical inverse Ising problem}\label{sec:model}

The dynamical inverse Ising problem, first studied in
 \cite{Roudi_Hertz_PRL_106_048702}, asks to reconstruct the couplings
and the time dependent external fields from output data of the real
system. These data are sets of configurations which can be considered as a snapshot of the
real system at different time steps. In our settings they are
generated according to the dynamics of the Ising model with real couplings.

Let's consider a system of $N$ nodes
connected by the couplings $J_{ij}$. The state of each node is represented by a spin
$\sigma$ that takes a discrete value in $\{-1,+1\}$.  In
eq.~\eqref{eq:prob}, we define a stochastic dynamics describing the
evolution of a configuration of spins at time $t$ to one at
time $t+1$. Here we consider the parallel update
dynamics for which, the state of all spins at time $t+1$ depends only
on the configuration at time $t$, and all spins evolve in parallel at
the same time.

\begin{equation}\label{eq:prob}
	P(\sigma_i(t+1)|\underline{\sigma}(t)) = \frac{ \exp[\beta \sigma_i(t+1) \sum_{j \neq i} J_{ji} \sigma_j(t)]}{2 \cosh(\beta \sum_{j\neq i} J_{ji} \sigma_j(t))}
\end{equation}

\noindent 
Therefore, the probability of finding a particular path (a trajectory $\underline{\sigma}(0), ..., \underline{\sigma}(T)$) 
with $T$ time-steps can be written as

\begin{eqnarray}
	W[\underline{\sigma}(T)|\underline{\sigma}(0)] & = & \prod_{t=0}^{T-1	}\prod_{i=0}^{N-1} \frac{\exp[\beta \sigma_i(t+1)\sum_{j\neq i} J_{ji} \sigma_j(t)]}{2 \cosh(\beta \sum_{j\neq i} J_{ji} \sigma_j(t)) } \label{eq:all_dyn} \\
	& = & \exp[\beta T \sum_i\sum_{j\neq i} J_{ji} \langle \sigma_i(t+1) \sigma_j(t) \rangle_T - \sum_{t,i}\log(2 \cosh(\beta\sum_{j \neq i} J_{ji} \sigma_j(t)) )]
\end{eqnarray}

\noindent where $\langle . \rangle_T$ means taking the average of the
data over the different times: $\langle \sigma_i(t) \rangle_T=T^{-1}
\sum_t \sigma_i(t)$.  In general we consider not only the case with
symmetric couplings where the detailed balance holds and the system
has an equilibrium distribution, but also the case with non-symmetric
couplings where no equilibrium distribution exists.

Note that, compared to the static inverse Ising problem where the
couplings are symmetric and the data are sampled from the Boltzmann
distribution, dynamical systems display a broader picture of phase
diagram such as limited cycles or non-equilibrium steady states and
thus obviously meet wider needs in real-world systems.

The log-likelihood function of this process is obtained by taking the
logarithm of the expression \eqref{eq:all_dyn}.  This function can be
maximized simply by a gradient descent method, using the derivative of
the log-likelihood with respect to the couplings. In the symmetric
case we obtain:

\begin{eqnarray}
	\frac{1}{\beta T} \frac{\partial \log W}{\partial J_{ji}} &=& \left\langle \sigma_i(t+1) \sigma_j(t) \right\rangle_T + \left\langle \sigma_j(t+1) \sigma_i(t) \right\rangle_T  - \left\langle \sigma_j(t) \tanh \left[\beta \sum_{k\neq j} J_{kj} \sigma_k(t) \right] \right\rangle_T - \left\langle \sigma_i(t) \tanh \left[\beta \sum_{k\neq i} J_{ki} \sigma_k(t) \right] \right\rangle_T \nonumber \\
	&=& c_{ij}^{\rm data}(t+1,t) + c_{ij}^{\rm data}(t,t+1) - c_{ij}^{J}(t+1,t)- c_{ij}^{J}(t,t+1),
\end{eqnarray}	

which implies a moment-matching condition for the local maximum of the likelihood function:
 
\begin{eqnarray}
  c^{\rm data}_{ij}(t+1,t) &=& \left\langle \sigma_i(t+1) \sigma_j(t) \right\rangle_T \\
  c_{ij}^{J}(t+1,t) &=& \left\langle s_j(t) \tanh \left[\beta \sum_{k\neq j} J_{kj} s_k(t) \right] \right\rangle_T.
\end{eqnarray}

In the non symmetric-coupling case, the equations take a simpler form:

\begin{equation}
  \frac{1}{\beta T}\frac{\partial W}{\partial J_{ji}} = c_{ij}^{\rm data}(t+1,t) - c_{ij}^{J}(t+1,t).
\end{equation}

So, starting from an initial condition, the couplings can be updated using the very general update rule:
$J_{ij}^{\rm new} = J_{ij}^{\rm old} + \eta ( c_{ij}^{\rm data}(t+1,t)
- c_{ij}^{J}(t+1,t))$, where $\eta$ is the learning rate. In the following we will mainly concentrate on symmetric
couplings but the generalization to non symmetric couplings is
straightforward. 

Note that, in the static case, the likelihood is a function of the
correlations and therefore is difficult to maximize. It is even difficult
to compute the likelihood exactly since the computational complexity
is exponential in the system size.  Here, as opposed to the static case,
the likelihood of the dynamical model can be easily computed exactly
with linear complexity in the number of samples and quadratic in the
system size.  Although several approximate methods have been studied
in the literature \cite{Roudi_Hertz_PRL_106_048702,
  Mezard_Sakellariou_jstat_2011, zhang12, Sakellariou_etal_arxiv_2011}, using mean-field approximations to accelerate the inference
process, in this paper we consider only exact likelihood-maximizing
methods which compute the correlations exactly given a set of
parameters.

Another difference between the dynamical and the static case comes
from how the data are taken (or produced).  In the static problem, the
data are made of spin configurations sampled independently from the
Boltzmann distribution. Thus the result does not depend on how the
data are acquired, only the number of configurations matters. At the
opposite, in the dynamical case, the configurations are generated
through a stochastic markovian process. Therefore it is possible to
control the way the data are produced by, for instance, making many
sets of data starting from different initial conditions. This liberty
is impossible in the static case and should in principle improve here
the inference process for systems where the dynamics gets trapped into
some region of the phase space.

Here we consider a collection of data parametrized by two variables:
the number of different trajectories $R$ and the number of time steps
$T$ of each trajectory. The total number of configurations is then
equal to $R\times T$. In this way, we collect the data under the form
of $R$ trajectories with length $T+1$ \cite{zhang12}, each of which
starts from $R$ different randomly chosen configurations. For each
trajectory we update $T$ times the system according to
eq.~\eqref{eq:prob}. The $R$ parameter let us tune the position from
which the dynamics began. It therefore allows to cover a larger part
of the configuration space. For example, when $R=1$, all
configurations are sampled starting from one initial configuration. In the
generated data, it will be useful to use $R>1$ in order to explore
better the phase space.

The $T$ parameter controls the number of time-steps of each trajectory. Again, in many cases, we gain more 
information on the system by considering many short trajectories instead of a long one. 
Now, when the system is ergodic, tuning the dynamics should not have
any influence on the results since the configuration can go from any
point of the configuration space to any other ones, which includes
somehow already the new random initial conditions. But if the
ergodicity does not hold, for example when the system is in a spin
glass phase, a dynamics starting from a particular configuration could
be trapped into a certain regime of the configuration space. Thus,
starting from different configurations helps exploring wider area of
the phase space and improves the data quality.

Let's now rewrite the likelihood function in terms of the parameters $R$
and $T$.  If we denote the sampled configurations by $\{\sigma^a(t)\}$,
with $a\in [1, R ]$ and $t\in [1, T+1]$, the likelihood function of
generating those configurations can be written as

\begin{eqnarray}\label{eq:likelihood}
	\mathcal{L}&=&\prod_{a=1}^{R}\prod_{t=1}^T W[\underline{\sigma}^{a}(t+1) | \underline{\sigma}^{a}(t)] \nonumber\\
	&=& \exp\left[\beta \sum_{i\neq j} J_{ji} \left\langle \langle \sigma_i(t+1) \sigma_j(t) \rangle_T \right\rangle_R - \sum_{a,t,i}\log(2 \cosh(\beta\sum_{j \neq i} J_{ji} \sigma_j^a(t)) )\right]
\end{eqnarray}

\indent where the notation $\langle . \rangle_R$ is used to represent the average over the different realizations.
Then, using a gradient descent, a set of couplings can be inferred by
maximizing the above likelihood.  As discussed previously, by maximizing
this likelihood, we will not only fit the data but also the noise in
the data. It will be particularly true in the cases where the dynamics gets trapped in
some small region of the phase space, or when the number of samples
$R \times T$ is small. Then, we will have difficulties in inferring the couplings as
well as recovering the topology of the graph.  In this work, in order to
span a large region of the temperature $\beta$, we choose $R=10$
and $T=500$. This is equivalent of taking $R=1$ and $T=5000$
in term of the number of configurations that are visited. By doing
this, however, we are exploring different regions of the phase space
and therefore we are able to perform a better reconstruction.

\section{Dynamic Inverse Ising problem on sparse graphs}\label{sec:sparse}

Most of the biological and social networks are sparse, meaning that
the average degree of those networks is finite ($c<<N$ ). To do
inference on sparse systems, it is very important to first determine
which edges are present, or equivalently which couplings are non-zero.
Therefore, when it is known as a prior that the underlying graph of
a system is sparse, the correct Bayesian inference process should take
into account this information under the form of a prior
distribution. Ignoring this prior information when maximizing the
likelihood will lead to over-fitting, as we discussed in
Sec.~\ref{sec:intro} and shown in Fig.~\ref{fig:maxlike} left.

\begin{figure}
	\includegraphics[width=0.6\textwidth]{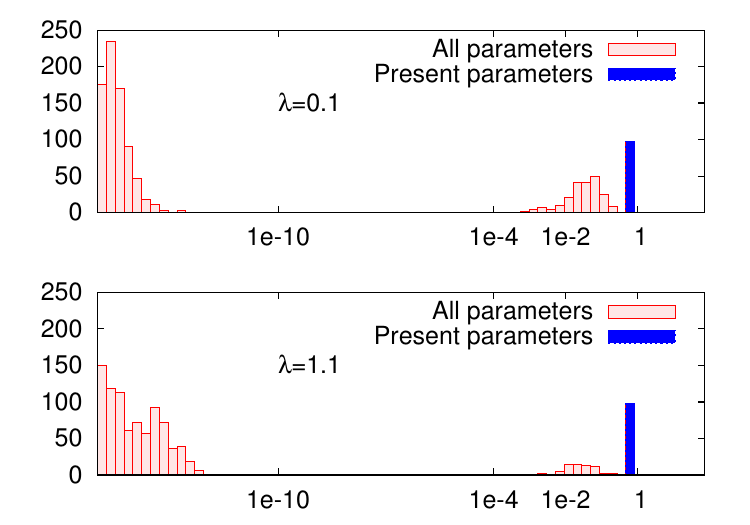}
	\caption{(Color online) Illustration of the results obtain by maximizing the likelihood together with the $\ell_1$ prior on a set of data generated by a parallel dynamics. The considered system is a 2D ferromagnet at low temperature, $\beta=0.61$ and we use $T=1000$ and $R=1$. The histograms show the parameters obtained for $\lambda=0.1$ (top) and $\lambda=1.1$ (bottom). Clearly the number of zero couplings change drastically when $\lambda$ changes.}
\label{fig_histo_lambda}
\end{figure}

The $\ell_1$-regularization is a well-known method to obtain sparse
solutions in inference problems. It has been widely used in compressed
sensing \cite{donoho2006compressed} and for the inverse Ising problems
\cite{wainwright2010AnnalHigh,tkavcik2014searching}. The $\ell_1$-based methods consist in
adding a Laplace prior: $e^{-\lambda \sum_{ij} |J_{ij}| }$ to the
system, in order to obtain a sparse solution. The major advantage of
the $\ell_1$ regularization is that, adding the Laplace prior does not
break the convexity of the likelihood function.
 this method is fully tractable, there are several drawbacks.
First, it is hard to fix the parameter $\lambda$ of the Laplace
distribution consistently.  When using a given value of $\lambda$,
many couplings will be put to zero in the maximization process. It is
possible to determine the maximum value of $\lambda$ analytically such
that, for $\lambda>\lambda_{\rm max}$, all the inferred couplings will
be zero. But then, the best value of $\lambda$ inside $[0,\lambda_{\rm max}]$ is
not given by the $\ell_1$ method. Indeed, by changing the value of
$\lambda$, the set of inferred zero-couplings will change accordingly
and it is therefore not possible to know what would be the best value
of $\lambda$.  As an example, in Fig. \ref{fig_histo_lambda}, we plot
the histogram of couplings inferred by maximizing the likelihood using
a $\ell_1$ prior and using two different values of $\lambda$. We can see
that, although it seems quite easy to choose the cut-off value on
whether couplings are zero or not, different values of $\lambda$ give a
different number of zero couplings and thus end with two different
topologies for the network.

Another drawback of the $\ell_1$ based-method is that the Laplace prior is
sometimes far from the distribution of the true couplings.
Thus the $\ell_1$ minimization will try to fit the couplings with a
Laplace distribution and lead to additional errors.  The Bayesian
prescription advises us that using the correct prior is always the
best choice to obtain the best performance, e.g.  in compressed
sensing problem \cite{PhysRevX.2.021005}, group testing problem
\cite{pan_group_testing} and error-correction problem
\cite{error_correction_ITW2013}.  However, in the inverse Ising problem,
we usually cannot use the correct prior distribution because otherwise
the problem would not be convex any more.

\section{Decimation Process}\label{sec:dec}

Decimation methods have been widely use in statistical mechanics in
various problems and it can be seen as a practical tool to reduce the
seach space of our problem. This idea is actually quite old
\cite{maslov}, especially in the context of optimization problems.  In
the context of the inverse Ising problem and more generally in machine
learning, this procedure can reduce greatly the overfitting over the
inferred parameters. It was therefore used to reduce the number of
parameters of the model to infer (see \cite{lecun1989optimal} for
method close to the one presented here). Another advantage of the
decimation procedure is that, by reducing the search space, the system
receives a feedback that it can exploit to solve the problem with more
accuracy. This is typically used when solving constraint satisfaction
problems using the marginals of the variables obtained by a
message-passing algorithm \cite{ricci2009cavity,mezard2002analytic}.



For the inverse Ising problem, the decimation process was introduced
to recover the topology of the Ising model and shown to provide a
significant improvement on the quality of the reconstruction over the
popular $\ell_1$-based methods \cite{decelle13}. In the static
problem, since the real likelihood function is hard to estimate
exactly, the decimation procedure is done by maximizing an
approximation of the likelihood function (the pseudo-likelihood
\cite{besag1975statistical}) and then by pruning a fraction of the
smallest couplings (in magnitude) by setting their values to
zero. This maximizing-pruning procedure is iterated until a given
criterion is reached.

In the dynamical problem, as we discussed in the last section, the
likelihood function can be easily estimated and thus the decimation
can be applied directly by using the true likelihood function. The
most intuitive way to fix the variables would be to set to zero one
variable at a time after having maximized the likelihood function over the remaining
parameters. However, in doing this way, the decimation
would be very slow (it would multiply the complexity of the algorithm by a factor $\log(N)$).
In order to make it fast without loosing accuracy, we can fix a finite
fraction of variables at each time step.  Doing so, the only parameter
in our decimation procedure is the fraction of couplings that are
pruned at each time-step.  A refined way to choose this fraction is,
at the first steps of the decimation, to prune a large amount of the
remaining parameters. Then, the fraction of couplings to decimate is
decreased gradually during the process.

We note that in the literature, another decimation approach has been
proposed \cite{lecun1989optimal,hassibi1993second} in a different
context.  In their approach, instead of pruning the parameters by
order of magnitude, the algorithm will in priority put to zero the
parameters that decrease the least the likelihood function. Let's
assume for instance that the maximum of the likelihood function has
been found. We can then make an expansion around this maximum ${J}^*$:

\begin{equation}
	\mathcal{L}({J}^*+{\delta J}) \sim \mathcal{L}({J}^*) + \frac{1}{2} \delta{J} H \delta{J},
\end{equation}

where $H$ is the Hessian of the likelihood function evaluated at
${J}^*$.  The goal is then to find a set of couplings in such a way
that putting them to zero will affect the least the likelihood
function. Therefore it asks to minimize $L= \frac{1}{2} \delta {J} H
\delta {J}$ such that $J_i+\delta J_i=0$ for the couplings that will
be pruned. This second-order decimation process will be denoted as
``Deci-O2'' in the following and will be compared with our approach.

The most important step of the algorithm is to decide when to stop the decimation. 
We use here the same criterion as in \cite{decelle13} which is defined by 
investigating the behaviour of the likelihood function when we decimate the parameters.
 
To describe its principle, let us note the set of non-zero/zero
couplings of the true system by $\bm{J_1}$/$\bm{J_0}$ respectively.
One would expect the following. When a coupling $J_{ij} \in \bm{J_0}$
is pruned during the decimation process, the likelihood should remain
constant as this parameter is not useful to describe the data. In
practice, and due to the over-fitting phenomena, the likelihood will
slightly decrease since the model is less accurate to fit the noise of
the data. Now, if a coupling $J_{ij} \in \bm{J_1}$ is pruned, this
parameter was necessary to describe correctly the data.  Therefore its
absence should impact the likelihood and we expect to observe its
value to decrease significantly. Hence, by looking at the
likelihood as a function of the number of the remaining parameters, we can
find the stopping point where the behaviour of the likelihood changes
significantly. That is, a discontinuity in the first derivative of
the likelihood function should be observed.

\begin{figure}[t]
   \begin{subfigure}[b]{0.45\textwidth}
	      \includegraphics[width=1\columnwidth]{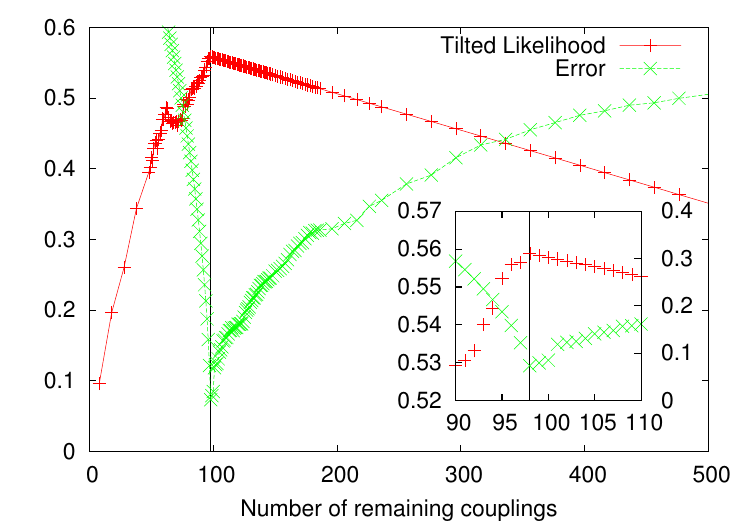}
		  \caption{}
	  \end{subfigure}
   \begin{subfigure}[b]{0.45\textwidth}
	   \includegraphics[width=1\columnwidth]{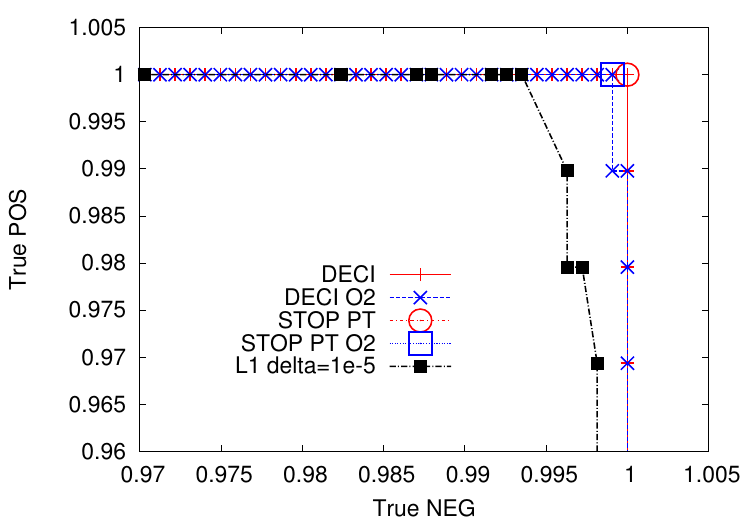}
		  \caption{}
	  \end{subfigure}
	\caption{(Color online) (a): Illustration of a typical Tilted Likelihood (TL) function. 
	In this case we are in a hard region, $N=49$, $\beta=1.5$ with $R=10$ and $T=500$ 
	of a 2D symmetric model with $J=\pm1$. 
	We observe a well-defined peak which indicates where to stop the decimation process.
	In the inset we show that the relative error is minimized at the peak. 
	In this case, the decimation process reconstructs perfectly the network 
	as indicated in the right panel.
	(b): the ROC curve for the same system in the left panel. 
	The ROC curves show the number of true positives on the y-axis and 
	true negatives on the x-axis. The goal is therefore to reach the 
	right-top angle for a perfect reconstruction. Here, the decimation process 
	and the Deci-O2 method are shown at different stages of the decimation 
	(at $T=0$ the curve start at (0,1)). 
	For the $\ell_1$ method, different points correspond to 
	different values of $\lambda$. 
	We can see that, our method works better than the two other ones. 
	The $\ell_1$ method is clearly less good even when considering the 
	best possible $\lambda$ ($\delta$ indicates the value of the cut-off but is chosen such that it goes into the gap separating zero couplings from the non zero ones (see Fig. \ref{fig_histo_lambda}).}
	\label{fig_tilted} 
\end{figure}

We will follow here the prescription of \cite{decelle13} in order to
characterized the stopping point for the decimation.  In order to
exhibit a peak at the point where to stop the decimation, the following
``tilted-Likelihood'' (TL) function has been considered

\begin{equation}
	\mathcal{L}^{\rm tilted}(x) = \mathcal{L}(x) -x \mathcal{L}_{\rm max} + (1-x)N \log(2),
\end{equation}

where $\mathcal{L}_{\rm max}$ is the maximum likelihood when all the
parameters are present (before decimating), and $x$ denotes the
fraction of couplings that remain unfixed. We also note
$\mathcal{L}(x)$ the maximum of the likelihood over the remaining
parameters (note that $\mathcal{L}(0)$ is the usual maximum likelihood
when all parameters are present).  The TL is constructed as
follows. We consider a new likelihood for a system of independent
variables but where its maximum value when all the couplings are
present ($x=0$) matches the likelihood of the system
$\mathcal{L}(0)=\mathcal{L}_{\rm indep}(0)$.  The likelihood of our
system of independent variables should decrease linearly (with $x$)
when decimating the parameters since they are here not useful to
describe the system. When $x=1$, both likelihoods will reach the value
$\mathcal{L}(1)=\mathcal{L}_{\rm indep}(1)=-\log(2)$ by
definition. The TL is constructed by looking at the difference between
the two likelihoods $\mathcal{L}_{\rm
  tilted}(x)=\mathcal{L}(x)-\mathcal{L}_{\rm indep}(x)$.  It can be
seen by this construction that the TL goes to zero when $x=1$ and
$x=0$. Therefore it should exhibit a maximum value between these two
limits.  When we start decimation, since we expect to decimate
couplings from $\bm{J_0}$ at the beginning, the likelihood of our
system will remain almost constant while the likelihood of the system
with independent variables will decrease more rapidly and make the TL
increases.  At a certain point, where we start to decimate parameters
that are present in the true system, the likelihood of our system will
decrease dramatically while the likelihood for the system of
independent variables will keep decreasing at a constant
rate. Therefore, a peak should appear at the point where the process
starts to decimate the wrong parameters.


\begin{figure}[t]
  \includegraphics[width=0.45\textwidth]{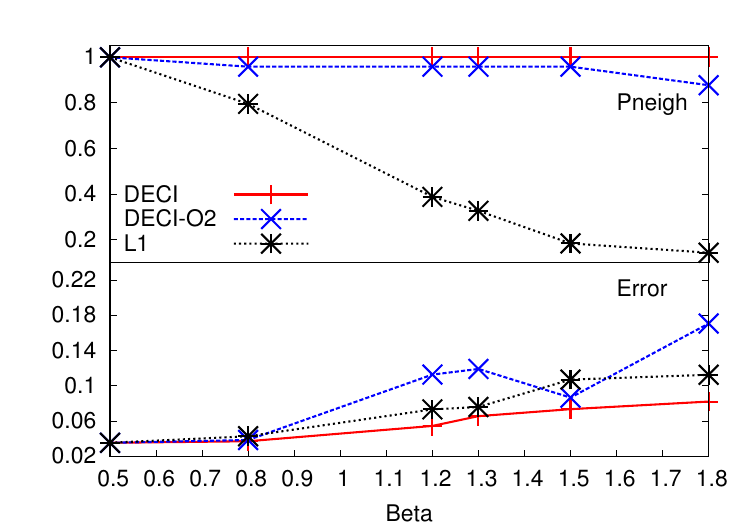}
  \caption{(Color online) Results obtained on a 2D symmetric model $J=\pm1$ with $N=49$, $R=10$ and $T=500$ for various values of $\beta$. On the top panel, we can see the topology reconstruction (eq. \ref{eq:pneigh}) of the three methods : Decimation, Deci-O2 and $\ell_1$. 
  We can see clearly that the decimation always recovers perfectly the topology of 
  the graph whereas $\ell_1$ performs poorly. The Deci-O2 method works quite close 
  to the decimation. On the bottom panel, we show the error on the reconstruction (eq. \ref{eq:diff}). All methods perform similarly with a slight advantage 
  towards the decimation. This behaviour can be explain by the fact that, often, the error in the topology reconstruction is dominated by the false negative couplings. The $\ell_1$  finds a lot of false positives but they are inferred quite close to zero (i.e. their true value) and therefore they do not contribute much to the reconstruction error.} \label{fig_ROC_GS}
\end{figure}

In order to illustrate the behavior of the TL in the whole decimation
process we take a symmetric 2D Ising model with $N=49$ spins and
$J=\pm1$ at $\beta=1.5$.  The decimation process is characterized by
the function $K(T)$ which defines how many couplings we put to zero at
each time-step.  Here we define $K(T)$ in such a way that for small
$T$ a large amount of couplings are
decimated. Then, we progressively diminish the number of decimated
couplings at each $T$ until the we end up decimating one coupling at each time-step, 
$K(T)=1$. This parametrization is quite flexible and we
did not observe significant changes as soon as $K(T)$ was small enough
after having decimated half of the system. We illustrate on
Fig. \ref{fig_tilted} the behavior of the TL as a function of the
remaining parameters.  We can see that the TL exhibits a maximum at
the point where the relative error, eq. \eqref{eq:diff}, is
minimized. On the right part of Fig. \ref{fig_tilted} we plot the ROC
curve which puts the number of true positives on the y-axis versus and
number of true negatives on the x-axis.  In Fig. \ref{fig_ROC_GS} we
compare the performance of our method with Deci-O2 and the $\ell_1$
method in recovering the topology.  The figure shows that our
decimation method stopped at a point where the topology is perfectly reconstructed
- making no false negative or false positive. The Deci-O2 works a bit worse
than our method, making one false negative at its stopping point. The $\ell_1$
method works the worst, making lots of false negatives and false
positives.

\begin{figure}[h!]
	\centering
   \begin{subfigure}[b]{0.45\textwidth}
	   \includegraphics[width=1\textwidth]{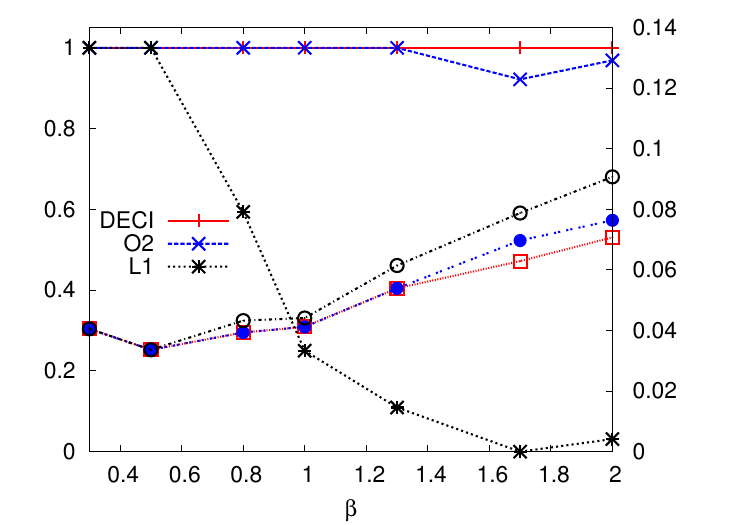}
		  \caption{}
	  \end{subfigure}
   \begin{subfigure}[b]{0.45\textwidth}
	   \includegraphics[width=1\columnwidth]{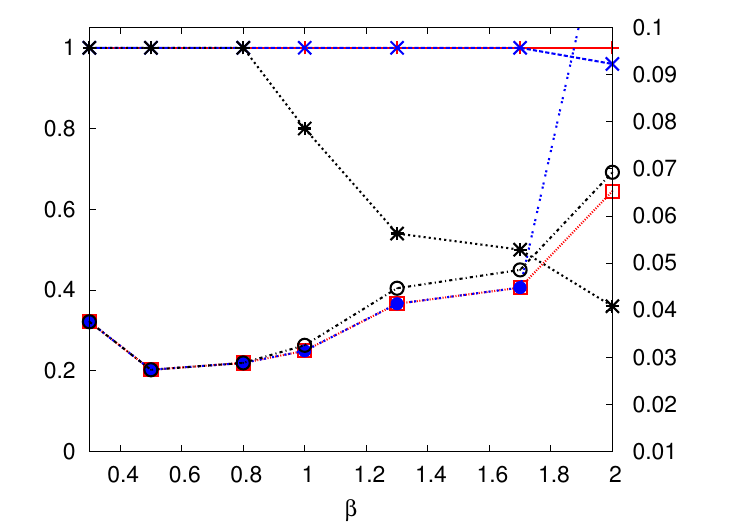}
		  \caption{}
	  \end{subfigure}

   \begin{subfigure}[b]{0.45\textwidth}
	   \includegraphics[width=1\columnwidth]{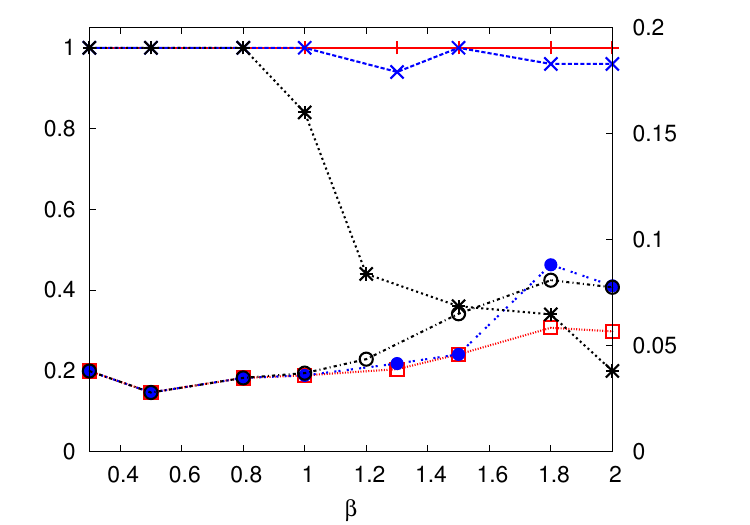}
		  \caption{}
	  \end{subfigure}
   \begin{subfigure}[b]{0.45\textwidth}
	   \includegraphics[width=1\columnwidth]{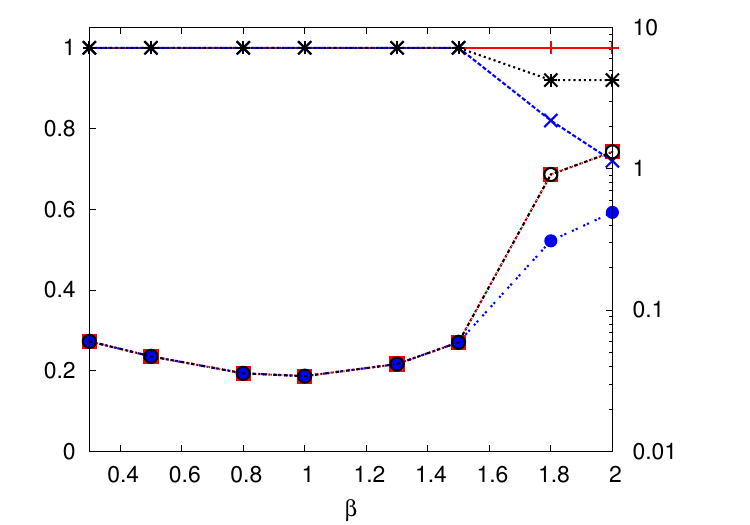}
		  \caption{}
	  \end{subfigure}
	\caption{(Color online) Comparison of the performance of the three methods on different topologies:
	(a) 3D lattice with $N=64$, $J=\pm1$, $R=10$ and $T=500$; 
	(b) ER random graph with $N=50$, $J=\pm1$, $R=10$ and $T=500$; 
	(c) Gaussian couplings on a random graph with $N=50$, 
	$\sigma_{\rm Gauss}=0.1$, $R=10$ and $T=500$; (d) Asymmetric couplings with 
	$J_{ij}=\pm1$ on a random graph with $N=50$, $R=1$ and $T=5000$.
In each figure the crosses correspond to the value of $P_{\rm neigh}$ 
(Eq.~\eqref{eq:diff}), the circles and the squares correspond to the 
relative error $\epsilon$ (Eq. \eqref{eq:diff}). 
From those figures we can conclude that the decimation works 
always much better than the $\ell_1$ method. The Deci-O2 method is comparable 
but always slightly worse than the decimation. 
The details concerning the comparison are described in the text.
}
    \label{fig_RES_3DRG}
\end{figure}

\section{Numerical results and performance comparison}\label{sec:result}
In this section we systematically study the performance of the
proposed method and we compared it with the $\ell_1$ method and the
Deci-O2 method on various topologies and with various parameters of
the kinetic Ising model.  These systems include $\pm J$ and
Gaussian-distributed couplings, on 2D lattices and on ER random
graphs. Concerning the decimation-based methods (decimation and
Deci-O2) at each time-step the likelihood is maximized over the
remaining couplings and a fraction of couplings is put to zero. Our
only free parameter was the fraction of decimated couplings at each
time-step. Concerning the $\ell_1$ method, the likelihood together
with the Laplace prior was maximized for a set of $\lambda$ chosen
inside $[0,\lambda_{\rm max}]$. Then, the value of $\lambda$ where the
ROC curve was the closest to the perfect reconstruction was chosen. We
should emphasize that this choice is highly non trivial and it would
not be possible to do the same in many real inference
problems. Finally, the topology is reconstructed by taking a cutoff
inside the gap separating the zero couplings from the others (see
Fig. \ref{fig_histo_lambda} for the gap). The error on the
reconstruction is computed over a new maximization of the likelihood,
this time without the Laplace prior but with the previously inferred
topology. This last step is done in order to not bias the value of the
inferred couplings because of the $\ell_1$ penalty term.

To evaluate the performance we consider two measures. The first
measure is the relative error of the reconstruction which is defined
as

\begin{equation}\label{eq:diff}
	\epsilon = \sqrt{ \frac{\sum_{i<j} \left(J_{ij}-J_{ij}^* \right)^2 }{\sum_{i<j} J_{ij}^2}  }.
\end{equation}

The second measure takes into account the reconstruction of the topology  
by counting the number of well-reconstructed neighborhoods. 
For each site $i$, the neighborhood of $i$ is said to be reconstructed if,
for all the couplings $J_{ij}$,  $\forall j\neq i$, the inference process has 
separated correctly the zero couplings from the present ones. 
We then sum over all the sites to obtain the measure

\begin{equation}
P_{\rm neigh}(\{J^*\}) = \frac{1}{N}\sum_{i=1}^{N} \delta\left(\sum_{\underset{J_{ij} \in \bm{J}_1}{j\neq i}}  \delta_{J^{*}_{ij},0} + \sum_{\underset{J_{ij} \in \bm{J}_0}{j\neq i}} (1-\delta_{J^{*}_{ij},0}) \right). \label{eq:pneigh}.
\end{equation}

This observable is zero if at least one error is made for all the
neighborhoods, and is one if all the neighborhoods are reconstructed
perfectly. The results are illustrated on Fig. \ref{fig_RES_3DRG}. We
can see from the figures that on various topologies and parameters,
the decimation algorithm clearly recovers the topology in all
cases. In comparison, the $\ell_1$ method does not work well to
recover the topology and usually finds slightly larger errors. However,
it is interesting to understand why the error in the $\ell_1$ method
performs as well as in the decimation and Deci-O2 cases while the
reconstructed topology is completely wrong. First of all, the
reconstructed topology is mainly affected by the fact that the
$\ell_1$ method is unable to find correctly who are the zero couplings
of the system. It means that, this method, even by tuning the value of
$\lambda$ (as we did here) is not close to the correct solution in
many cases. But, as the topology is wrong mainly because of the false
positives, the inferred value of these couplings will remain close to
zero and therefore they do not contribute significantly to the
relative error. Finally it is also interesting to see that the Deci-O2
method has a behaviour very similar to the one of the decimation while
always slightly worse than our method, especially in the topology
reconstruction.


\section{Conclusion and discussion}\label{sec:con}
In this paper we proposed a simple method based on a decimation
process to the inference of the dynamical inverse Ising problem on
sparse graphs.  We claim that our method is a much more robust
approach than the maximum-likelihood method, which is prone to
overfitting. In addition, unlike the $\ell_1$ approach, we do not
assume an artificial prior on the couplings, and we do not have any
parameter to tune.  Experimental results show that our method
outperforms $\ell_1$-based method on various topologies and with
different coupling distributions. It also shows improvement over a
decimation based method using the second derivative of the likelihood
function.

Including our method, many existing approaches to the inverse Ising
problems are based on a ``maximum \`a posterior'' method. However, in
cases where the amount of data is small or when the data quality is
low, it would be very helpful to develop a real Bayesian inference
method considering the average over all possible couplings and not
focusing on the value of the couplings that maximize the posterior
distribution. We leave this for future work.

Though all the study in this paper is theoretical, we note that it would 
be very interesting to test our method on real-world problems, probably with
missing or noisy data. We leave this for future work.

\begin{acknowledgments}
P.Z. has been supported by AFOSR and DARPA under the grant FA9550-12-1-0432. A.D. has been supported by the FIRB project n. RBFR086NN1.
We thank anonymous reviewer for pointing out work of 
S. Yu. Maslov and reference \cite{maslov}
\end{acknowledgments}

\bibliographystyle{unsrt}
\bibliography{Bib}

\end{document}